\documentclass[%
 reprint,
superscriptaddress,
 amsmath,amssymb,
 aps,
 showkeys,
]{revtex4-2}

\usepackage{graphicx}
\usepackage{dcolumn}
\usepackage{bm}
\usepackage{xcolor}
\usepackage{tikz}
\usepackage[english]{babel}
\usepackage{cleveref}
\usepackage{todonotes}

\makeatletter
\let\ORIbbl@fixname\bbl@fixname
\def\bbl@fixname#1{%
  \@ifundefined{languagealias@\expandafter\string#1}
    {\ORIbbl@fixname#1}
    {\edef\languagename{\@nameuse{languagealias@#1}}}%
}
\newcommand{\definelanguagealias}[2]{%
  \@namedef{languagealias@#1}{#2}%
}
\makeatother

\definelanguagealias{en}{english}
\definelanguagealias{EN}{english}

\begin{document}

\preprint{APS/123-QED}

\title{Hydroelastomers: soft, tough, highly swelling composites}

\author{Simon Moser}
\affiliation{Department of Materials, ETH Z\"{u}rich, Switzerland}
\author{Yanxia Feng}%
\affiliation{Department of Materials, ETH Z\"{u}rich, Switzerland}
 
 \author{Oncay Yasa}
 \affiliation{Department of Mechanical and Process Engineering, ETH Z\"{u}rich, Switzerland}
 
 \author{Stefanie Heyden}%
\affiliation{Department of Materials, ETH Z\"{u}rich, Switzerland}

\author{Michael Kessler}%
\affiliation{Institute of Materials, EPFL, Switzerland}
 
 \author{Esther Amstad}%
\affiliation{Institute of Materials, EPFL, Switzerland}

 \author{Eric R. Dufresne}%
\affiliation{Department of Materials, ETH Z\"{u}rich, Switzerland}

 \author{Robert K. Katzschmann}
 \email{rkk@ethz.ch}
 \affiliation{Department of Mechanical and Process Engineering, ETH Z\"{u}rich, Switzerland}

 \author{Robert W. Style}%
 \email{robert.style@mat.ethz.ch}
\affiliation{Department of Materials, ETH Z\"{u}rich, Switzerland}





\date{\today}

\begin{abstract}
Inspired by the cellular design of plant tissue, we present a new approach to make versatile, tough, highly water-swelling composites.
We embed highly swelling hydrogel particles inside tough, water-permeable, elastomeric matrices.
The resulting composites, which we call \emph{hydroelastomers}, show little softening as they swell, and have excellent fracture properties that match those of the best-performing, tough hydrogels.
Our composites are straightforward to fabricate, based on commercial materials, and can easily be molded or extruded to form shapes with complex swelling geometries.
Furthermore, there is a large design space available for making hydroelastomers, since one can use any hydrogel as the dispersed phase in the composite, including hydrogels with stimuli-responsiveness.
These features should make hydroelastomers excellent candidates for use in soft robotics and swelling-based actuation, or as shape-morphing materials, while also being useful as hydrogel replacements in a wide range of other fields.

\end{abstract}
\maketitle


\section{Introduction}

Hydrogels are perhaps the most important and diverse form of soft material.
In biology, they play a vital role as a main component of living tissue.
Elsewhere, hydrogels have found widespread usage across diverse disciplines including tissue engineering~\cite{saroia_review_2018}, drug delivery~\cite{li_designing_2016}, food science~\cite{williams_chapter_2021}, agriculture~\cite{qu_chitosan-based_2020}, and soft robotics~\cite{baumgartner_resilient_2020}.
This broad range of uses is tied to their common physical properties.
Hydrogels' high water content means that they are often biocompatible~\cite{saroia_review_2018}, and able to host biochemical reactions.
They can swell in volume by absorbing hundreds of times their weight in water, allowing actuation and large changes in structure. 
Furthermore, hydrogels can be made stimuli-responsive: capable of changing material properties or degrading in response to external stimuli like light, pH, temperature, or electric fields~\cite{koetting_stimulus-responsive_2015}.

However, there are certain challenges to working with hydrogels.
Hydrogels usually have properties that can vary by orders of magnitude as they swell from tough, stiff, dry, glassy solids to brittle, soft, highly-swollen gels~\cite{delavoipiere_poroelastic_2016}.
They also dehydrate quickly in air~\cite{bai_transparent_2014}, and can be tricky to adhere to most surfaces~\cite{yang_hydrogel_2020}.
Perhaps the greatest challenge is that simple hydrogels are very brittle.
To counter this shortcoming, recent advances have developed a number of strategies to significantly improve toughness~\cite{gong_why_2010,sun_highly_2012,lin_muscle-like_2019,sun_physical_2013,kim_fracture_2021,li_stiff_2014,hubbard_hydrogelelastomer_2019}.
However, we currently lack an easy strategy to prepare materials that combine the useful features of hydrogels, such as stimuli-responsiveness, with the toughness of biological tissue or commercial elastomers.

\begin{figure}
    \centering
    \includegraphics[width=\columnwidth]{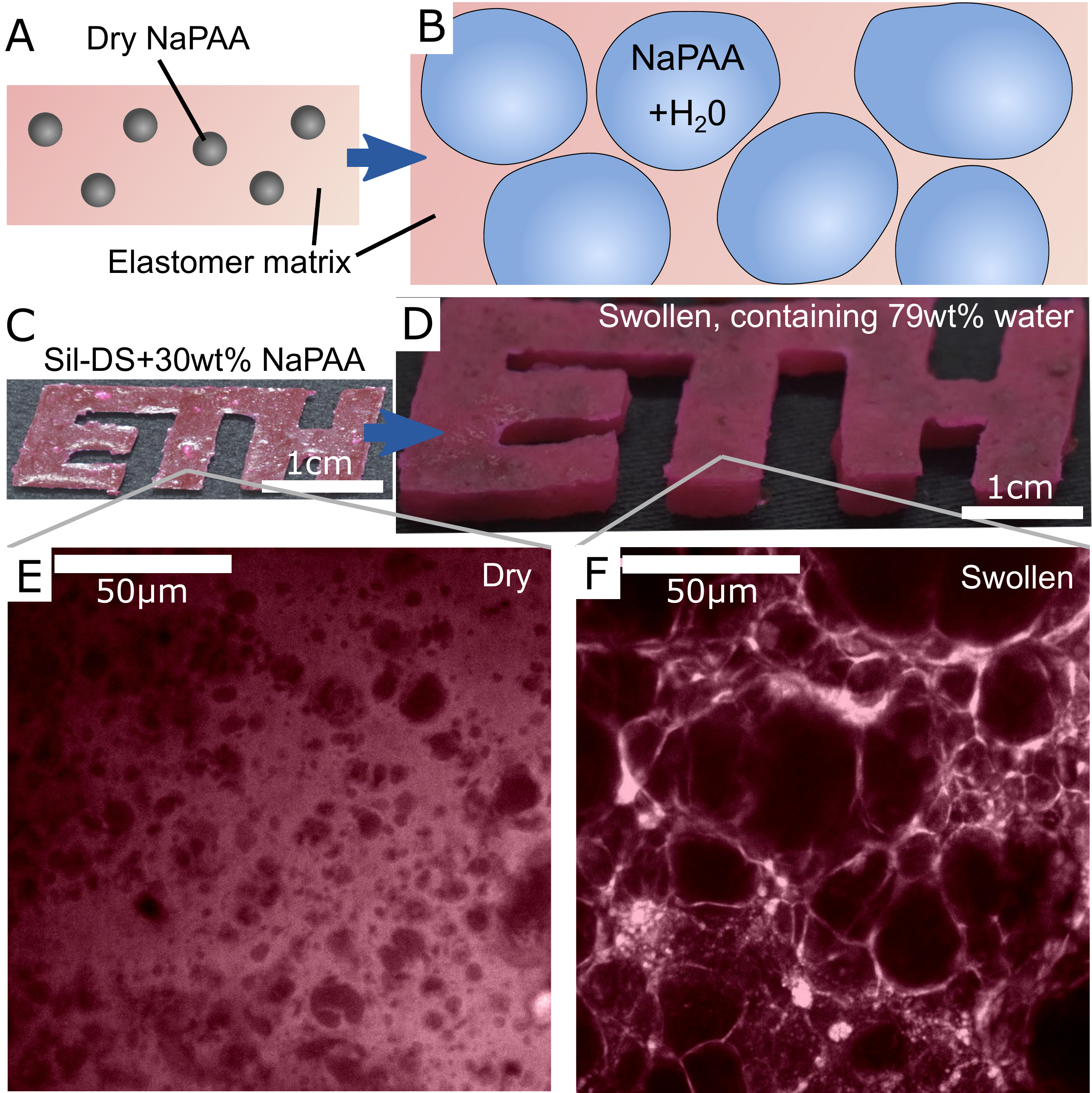} 
    \caption{An overview of hydroelastomers. (A,B) show the swelling principle of the material, whereby highly swelling microgels are embedded in tough, permeable elastomers to make tough, swellable composites. (C,D) show a sample made of 30\% sodium polyacrylate (NaPAA) in dragon-skin silicone before and after swelling in water. The color comes from rhodamine B dye. (E,F) Confocal microscopy images showing the microstructure of the samples in C, D. Dark areas are the hydrogel.}
    \label{fig:intro}
\end{figure}

To address this issue, we introduce a type of highly water-swelling material, which we call a \emph{hydroelastomer}.
This consists of microscopic hydrogel particles (microgels) in tough, water-permeable, elastomeric matrices (\Cref{fig:intro}).
Our material design is motivated by nature's strategy of combining highly swellable osmotic inclusions (cells) inside tough matrices to produce tough materials like plant tissue~\cite{kataruka_swelling_2021}.
Our strategy also builds on previous work showing that liquid inclusions can enhance fracture and stiffness properties of elastomeric materials~\cite{style_stiffening_2015,kazem_extreme_2018,style_solidliquid_2021,kataruka_swelling_2021}.
The resulting materials show fracture properties that compete with (and can actually outperform) the toughest available hydrogels, while their stiffness is surprisingly constant as they swell.
The preparation method is not oxygen-sensitive (unlike many hydrogel syntheses~\cite{simic_oxygen_2021}), and is based on materials such as silicones and polyurethanes -- common sealants with good adhesive properties.
We show that the materials are highly adaptable, as they can be  straightforwardly formed into complex geometries with non-uniform structures.
Hence, we envisage that hydroelastomers should have a range of potential uses replacing hydrogels or other soft materials in applications including soft robotics~\cite{baumgartner_resilient_2020,marchese_recipe_2015}, shape-morphing materials~\cite{sydney_gladman_biomimetic_2016}, swelling sealants~\cite{kleverlaan_deployment_2005}, and water-retention in agriculture~\cite{qu_chitosan-based_2020}.

\section{Material Creation}

To create our composites, we embed dry sodium polyacrylate (NaPAA) microgels in a range of soft elastomers, and then swell them in water (\Cref{fig:intro}A,B).
In principle, we could use any of the wide range of hydrogels as microgels for this purpose.
However, NaPAA is convenient as it is a common, commercial, super-absorbent polymer, capable of absorbing hundreds of times its weight in water.
We fabricate dry NaPAA powder via emulsion polymerization, to produce particles with an approximate diameter of 30~$\mu$m when fully swollen in water (see Materials and Methods and Supplement for further details). 
For the elastomeric matrix, we use common, commercial elastomers: Smooth-on Dragonskin 30 (Sil-DS), a relatively stiff, tough silicone, with a measured Young's modulus, $E=1.12$~MPa Smooth-on Ecoflex 10 (Sil-EC), a much softer silicone, with $E= 33$ kPa; and Smooth-on Vytaflex 40 (PU-VY) a polyurethane, with $E=1.4$ MPa.
These all swell very little in water~\cite{bian_rediscovering_2021}, but are permeable to water transport~\cite{lee_solvent_2003}.

We note that almost any water-permeable elastomer or gel could be used as a matrix material.
Indeed, one could use tough hydrogels -- which would result in materials that are conceptually similar to microgel-reinforced hydrogels~\cite{hu_microgel-reinforced_2011,hirsch_3d_2021}.
The only constraint on the matrix is that it must be soft enough to be deformed by the osmotic pressures generated in the swelling microgels.
As shown in the Supplementary Materials, using data from \cite{hamdan_draw_2015}, this osmotic pressure can reach up to almost 100 MPa.
Thus, any material with $E\lesssim 10$MPa, would be appropriate.

The final fabrication step involves mixing dry NaPAA powder with the liquid precursors of the matrix material with an centrifugal mixer.
The mixture is then degassed and cured overnight in an oven at 40$^\circ$C.
We use a mass fraction, $\phi_p$, of dry microgel in the as-prepared composite in the range of 10-30wt\%. 
It is difficult to work with higher fractions without trapping air bubbles, which compromise the material properties.
This is especially true when working with matrix precursors with a large viscosity.

A typical example of a composite made from 30wt\% of dry NaPAA in Sil-DS is shown in \Cref{fig:intro}C,D.
After swelling in de-ionized water, the mass of the sample increases by 275\% until it reaches an equilibrium swelling that is almost 80wt\% water.
The microgels expand significantly upon swelling, ultimately stretching the water-permeable matrix to form thin walls between the individual inclusions (\Cref{fig:intro}E,F).

\section{Stiffness}
\begin{figure}
    \centering
    \includegraphics[width=\columnwidth]{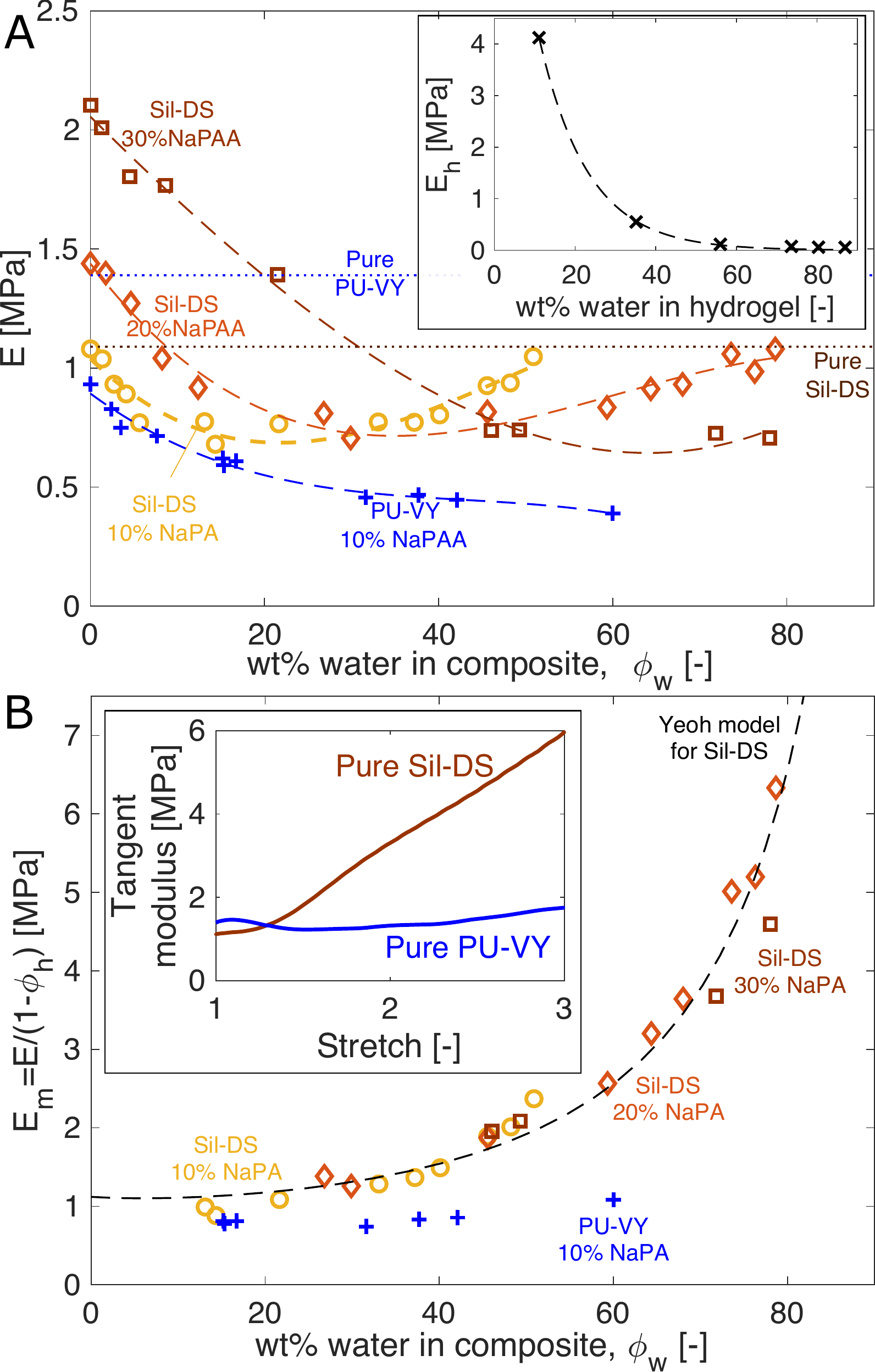} 
    \caption{Composite stiffness is a surprisingly weak function of swelling. (A) Composite stiffness versus water content. The yellow, orange, and brown data sets are Sil-DS composites with 10, 20, and 30wt\% of dry NaPAA, respectively. The blue data set is a PU-VY composite with 10wt\% of dry NaPAA. The dashed curves are best-fit cubic polynomials, while the dotted lines show $E$ for pure Sil-DS and PU-VY. Inset: the stiffness of the NaPAA (measured in bulk samples) as a function of its water content. The dashed curve is a best-fit exponential function. (B) The effective modulus of the matrix versus swelling. We use only data points from A, where the microgels contain more than 50wt\% water, to give $E_m=E/(1-\phi_h)$. The dashed line is our theoretical prediction of $E_m$ for Sil-DS with 10wt\% of dry NaPAA. The inset shows the tangent moduli of pure, dry Sil-DS and pure, dry PU-VY when stretched uniaxially. These are qualitatively similar to the data in the main figure.}
     \label{fig:stiffness}
\end{figure}

We study the mechanical properties of the composites as a function of their water content.
\Cref{fig:stiffness}A shows the Young's modulus of pure NaPAA hydrogel (inset), and silicone and polyurethane composites as a function of the total mass fraction of water in the samples, $\phi_w$.
The silicone composites are initially formulated with 10, 20, and 30wt\% of dry NaPAA, while all polyurethane composites are made with 10wt\% of dry NaPAA.
The dry silicone composites are stiffer than their respective pure matrix materials, while the polyurethane composites are softer.
Upon initial swelling, all the samples start to soften.
However, interestingly, the silicone-based ones subsequently stiffen upon further swelling -- especially the 10 and 20wt\% samples.

We can understand the qualitative changes in stiffness during swelling in terms of the properties of the microgels and the surrounding matrix.
When dry, NaPAA is a glassy solid that is orders of magnitude stiffer than the surrounding matrix, and thus would be expected to stiffen the composite.
This probably explains why the dry, higher NaPAA-content silicone composites are initially stiffer than pure silicone (in contrast, residual water in the `dry' NaPAA powder probably hinders polymerization of the polyurethane samples, reducing their stiffness).
As the composites swell, their stiffness is determined by a competition between softening of the microgels and stretching of the matrix.
The microgel softening occurs rapidly as it swells, as shown by the inset in \Cref{fig:stiffness}A.
This is likely responsible for the initial decrease in stiffness of all the composites upon swelling.
By contrast, changes in composite stiffness at higher swellings are probably controlled by stretching of matrix phase.
This conclusion is supported by the data in the inset of \Cref{fig:stiffness}A, which shows that microgels with a water content of $>50$\% have a stiffness that is essentially negligible in comparison to that of the composite.

We gain further insight into the origin of the large-swelling behavior by isolating the contribution of the matrix for the data in \Cref{fig:stiffness}A.
The law of mixtures estimates composite stiffness as $E = E_m (1-\varphi_{h})+E_{h} \varphi_{h}$, where $E_m$ is the effective modulus of the matrix, $E_{h}$ is the microgel stiffness (the subscript $h$ stands for hydrogel), and $\varphi_{h}$ is the volume fraction of swollen microgels in the composite.
For the highly swollen composites, where $E_h\approx0$, we can estimate $E_m=E/(1-\phi_h)$ (we assume $\varphi_h=\phi_h$, the mass fraction of hydrogel).
Plotting $E_m$ for the samples where the microgels are more than 50\% water collapses the silicone data nicely (\Cref{fig:stiffness}B).
This analysis highlights that the silicone matrix appears to stiffen dramatically as it is stretched by swelling -- up to 6 times it's original modulus.
By contrast, the polyurethane composite shows no real evidence of strain stiffening.
This is consistent with the strain-stiffening behavior of the pure matrix materials (shown for uniaxial tension in the inset of \Cref{fig:stiffness}B).

Indeed, we can use the measured nonlinear properties of the matrix materials to predict $E_m$ with good accuracy.
We fit the results of a uniaxial tension experiment on a pure Sil-DS sample with a hyperelastic Yeoh model.
Then, we create a model of the composite as a sphere of this matrix material containing a growing spherical cavity.
We inflate the cavity step-wise.
At each step, we calculate the average incremental modulus of the matrix in response to a unidirectional stretch, while holding the cavity volume fixed.
The results for a silicone composite with 10wt\% of dry NaPAA are given as the dashed curve in \Cref{fig:stiffness}B (further curves for higher NaPAA loadings, and details of the calculation are given in the Supplementary Materials).
The model captures the trend shown by the data, suggesting that the increasing stiffness of the composite at high swelling is indeed caused by strain stiffening of the matrix.

\section{Fracture Energy}

\begin{figure}
    \centering
    \includegraphics[width=\columnwidth]{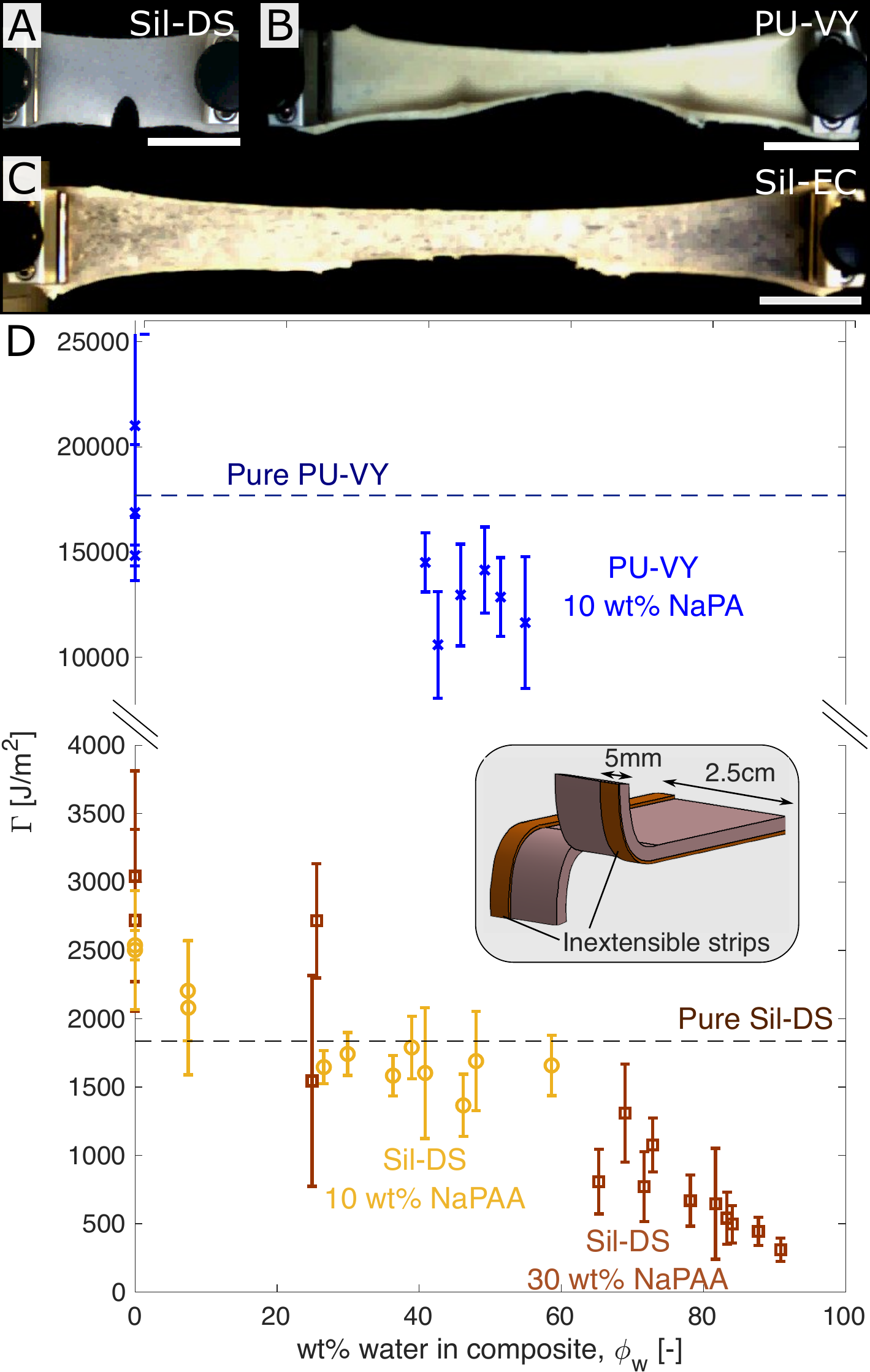} 
    \caption{The fracture strength of swelling composites. (A-C) Examples of mode I cracks at the point of failure for swollen composites made from different matrix materials. The samples are made with 10~wt\% of dry NaPAA and are swollen to $\phi_w=30,51,60$~wt\% respectively. All scalebars are 2.5~cm long. (D) Fracture energy versus swelling for polyurethane (blue) and silicone (yellow/brown) composites. The inset shows a schematic of the test geometry used. Flexible, inextensible strips are glued to the sample edges to prevent leg stretching during the test. The sample thickness depends on the swelling state of composite.}
    \label{fig:fracture}
\end{figure}

The composites inherit much of the toughness of their corresponding matrix material, even when highly swollen with water.
This fact is demonstrated in \Cref{fig:fracture}A-C, which show examples of notch tests on composites with different matrix materials at different water contents (all originally made with 10wt\% of dry NaPAA).
In each image, the sample is at the point of crack propagation.
More quantitatively, \Cref{fig:fracture}D shows how fracture energy varies with swelling for different composites, measured with trouser tests (\emph{c.f.} the schematic \cite{long_fracture_2016,huang_energy-dissipative_2017}).
Interestingly, for dry composites, $\Gamma$ is actually larger than that of the pure matrix material, especially for the silicone composites.
However, this is not surprising, as it is known that stiff, glassy microparticles act to toughen silicones and polyurethanes~\cite{paul_fillers_2010,petrovic_structure_2000}.
As the composites swell, $\Gamma$ reduces.
In particular, for the silicone composites, we see an almost linear decrease with $\phi_w$, approaching 0 as $\phi_w\rightarrow 1$.

The observed linear decrease in $\Gamma$ with swelling conceptually fits with a simple law of mixtures approach: $\Gamma=\Gamma_m (1-\phi_h)+\Gamma_h \phi_h$, where now $\Gamma_m$ and $\Gamma_h$ are the matrix and microgel fracture energies, respectively.
The microgels should be very brittle, so $\Gamma_h \ll \Gamma_m$, and $\Gamma\approx\Gamma_m(1-\phi_h)$.
When $\Gamma_m$ is not a strong function of stretch, this expression yields a linear drop off in fracture energy with $\phi_h$, as seen in \Cref{fig:fracture}A.
We note, however, that a stretch-independent value of $\Gamma_m$ is rather unexpected, as recent experiments have shown that stretching silicone can have a significant effect on its fracture energy~\cite{lee_sideways_2019}.

\section{Applications}

\begin{figure}
    \centering
    \includegraphics[width=\columnwidth]{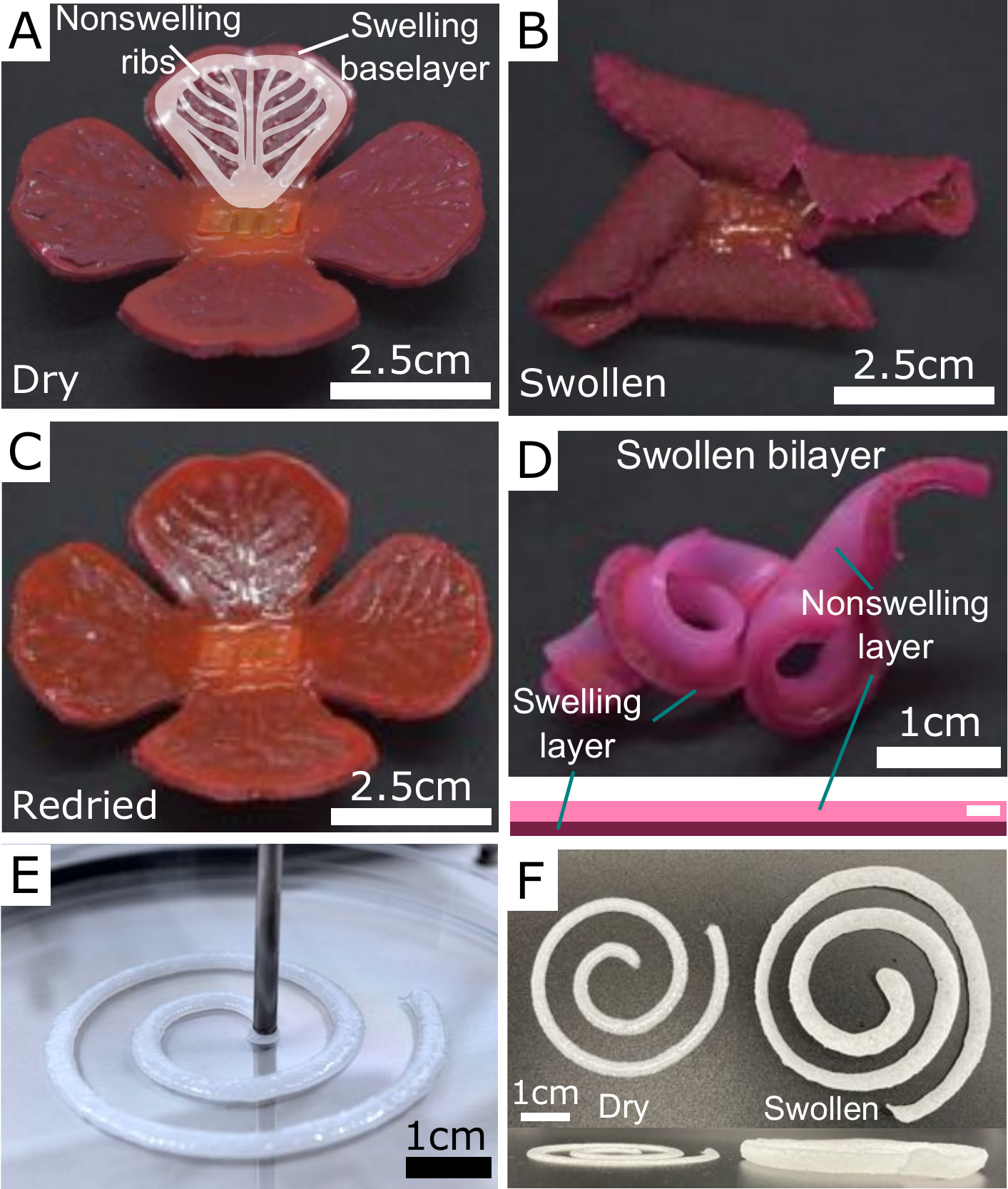} 
    \caption{The fabrication process allows easy creation of complex swelling geometries. (A-C) A flower created with a 1mm-thick swelling layer of 10wt\% NaPAA in Sil-DS, bonded to a 1mm-thick rib structure, as shown in the schematic. The color comes from rhodamine B and Sudan I dye. After immersing in water, the petals curl up, before returning to the original shape when dried out. The slight color change is due to dye leaching during swelling. (D) A swollen bilayer consisting of a strip of non-swelling Sil-DS (dyed light pink) bonded to a swelling strip of swellable composite (dyed dark pink). (E-F) The curing composite is shear thinning, allowing it to be printed by extrusion.}
    \label{fig:apps}
\end{figure}

The simplicity of the fabrication process gives great flexibility in terms of creating objects with complex swelling characteristics.
\Cref{fig:apps} shows a few simple demonstrations of using hydroelastomers as a swelling material.
For example, \Cref{fig:apps}A shows a flower where the `petals' are made of a 1 mm-thick layer of swelling material (Sil-DS with 30wt\% of dry NaPAA) underlying another 1 mm-thick, non-swelling rib structure, made with pure Sil-DS (\emph{c.f.} schematic).
We use a two-step molding process.
When swollen in water, the petals curl upwards due to differential swelling, before repeatably returning to their original state when dried out.
Similarly, \Cref{fig:apps}B shows the results of swelling a simple, 15 cm-long bilayer consisting of a strip of swelling material (dark pink), bonded to non-swelling strip of pure Sil-DS (light pink).
Upon swelling, the material drastically changes its form by curling up into a gut-like shape~\cite{savin_growth_2011}.
Videos of the flower and bilayer swelling are given in the Supplementary Materials.
These illustrate the timescale for swelling of these materials.

Beyond the usage of molds, the composites can also be printed, as the curing microgel/polymer mixture has suitable shear-thinning characteristics.
\Cref{fig:apps}C shows a simple spiral of printed Sil-DS with 10wt\% NaPAA during printing and after swelling.
In future, we anticipate that complex 3D shapes can be created by directly mixing particles and polymer in the printer to continuously tune particle concentration during printing.
This would allow the simple manufacture of complex swelling morphologies.

\section{Discussion \& Conclusions}

\begin{figure}
    \centering
    \includegraphics[width=\columnwidth]{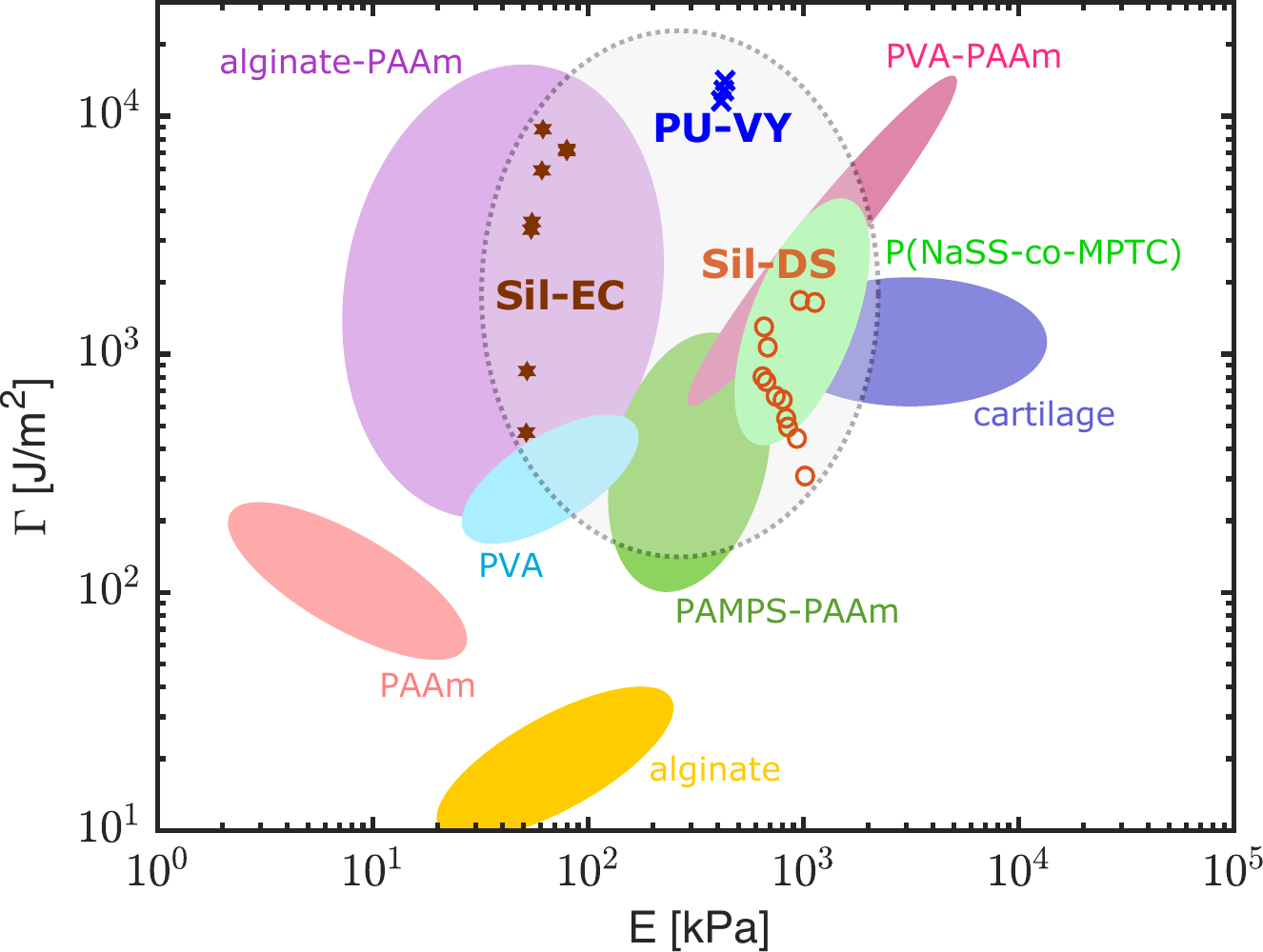} 
    \caption{An Ashby diagram showing fracture energy and stiffness that compares our composites with state-of-the-art, water-swellable materials, modified from~\cite{li_stiff_2014}.
    Out of our composites, only those with $\phi_w>48\%$ are shown here.
    The data from \Cref{fig:stiffness,fig:fracture} are combined by using the best fit curves in \Cref{fig:stiffness} to calculate the corresponding $E$ for each point in \Cref{fig:fracture}. PAAm: polyacrylamide, PVA: poly(vinyl alcohol), PAMPS: poly(2-acrylamido-2-methylpropanesulfonic acid), P(NaSS-co-MPTC): sodium p-styrenesulphonate-co-3-(methacryloylamino)propyl-trimethylammonium. Original references: PAAm, alginate, alginate-PAAm~\cite{sun_highly_2012}, PVA~\cite{zhang_anisotropic_2012}, PAMPS-PAAm~\cite{gong_why_2010}, PVA-PAAm~\cite{li_stiff_2014}, P(NaSS-co-MPTC)~\cite{sun_physical_2013}, cartilage~\cite{wegst_mechanical_2004}.}
    \label{fig:ashby}
\end{figure}

Our results show that our composites are very competitive when compared to other, tough, swelling materials.
\Cref{fig:ashby} shows an Ashby diagram, modified from~\cite{li_stiff_2014}, showing the mechanical properties of a selection of state-of-the-art, water-swellable materials.
Data points for our composites are shown for materials swollen with $>48\mathrm{wt}\%$ water. 
Our composites are in a similar range of stiffness to hydrogels, while having fracture properties that compare to state-of-the-art synthetic hydrogels.
The polyurethane composites perform particularly well: matching, if not exceeding the best hydrogels available.

Our composites also have some significant advantages over the hydrogels in \Cref{fig:ashby}.
For instance, the matrix materials, \emph{i.e.} silicones and polyurethanes, are robust, commercial materials that can be used in harsh conditions.
Silicones in particular have good heat resistance, chemical stability, and weatherability \cite{shit_review_2013}.
We anticipate that these characteristics will carry over to the composites.
Our hydroelastomers are simple to prepare.
Once the microgels are created via a straightforward emulsion polymerization, our composites are fabricated by mixing the ingredients together, briefly degassing, and then shaping with a mold, or via extrusion.
In contrast, chemically crosslinked gels typically require oxygen-free conditions~\cite{simic_oxygen_2021}, polyvinyl alcohol gels require freeze-thaw~\cite{stauffer_polyvinyl_1992} or annealing and re-swelling~\cite{li_stiff_2014}, and polyampholyte and double network gels require dialysis or re-swelling steps~\cite{sun_physical_2013,nakajima_double-network_2021}.
Furthermore, the mechanical properties of our composites vary surprisingly little as they swell.
This can be contrasted with hydrogels, which often go through a glass transition as they dry out (\emph{e.g.} \cite{delavoipiere_poroelastic_2016})

In conclusion, we present a simple, highly water-swellable class of material with a toughness that matches state-of-the-art tough hydrogels, while being based on a completely different microstructural architecture.
These hydroelastomers are easy to fabricate, and ideal for creating complex, 3D-printed, swelling materials.
We anticipate this material being of great use in swelling-controlled actuation~\cite{ionov_hydrogel-based_2014}, and shape-morphing materials~\cite{sydney_gladman_biomimetic_2016}.
The wide range of choices of matrix and inclusion materials gives great design flexibility in controlling composite properties.
In particular, the material allows us to combine the features of hydrogels with desired material properties (\emph{e.g.}, stimuli-responsiveness, or high swelling capacity) with a tough elastomeric matrix
This strategy could be used to make tough composites that swell in response to triggers such as temperature, light, or electromagnetic fields~\cite{koetting_stimulus-responsive_2015}.
Currently, the main limitation on the swelling process is the speed of permeation of water through the matrix~\cite{bian_rediscovering_2021} (though this does aid in hindering dehydration).
In the future, we anticipate substantially speeding up the swelling process by using matrix materials with higher water permeability, such as tough silicone hydrogels~\cite{musgrave_contact_2019,si_silicone-based_2016}.

\begin{acknowledgments}
This work was partially funded through the Swiss National Science Foundation through the NCCR Bio-inspired Materials.
Author contributions: S.M., O.Y., E.R.D., R.K.K. and R.W.S. designed and supervised the research. M.K. and E. A. developed the microgels and performed the 3D printing experiments. S.M. and Y.F. performed all other experiments. S.H. developed the theoretical model for stiffening. S.M., Y.F., R.K.K. and R.W.S. analyzed the data, S.M., O.Y., E.R.D., R.K.K. and R.W.S. wrote the paper.
\end{acknowledgments}

%


\end{document}